# How average is average? Temporal patterns in human behaviour as measured by mobile phone data – or why chose Thursdays.

Marina Toger [a,c]*, Ian Shuttleworth [b,c], John Östh [a,c].

[a] Department of Social and Economic Geography, Uppsala University, Box 513, SE-751 20 Uppsala, Sweden;
[b] School of Natural and Built Environment, Queen's University Belfast, Belfast BT7 1NN;
[c] CALISTA centre for applied spatial analysis, Uppsala University, Sweden.

* Corresponding author: Marina Toger, Department of Social and Economic Geography, Uppsala University, Box 513, SE-751 20 Uppsala, Sweden; 2020-04-26. Email: marina.toger@kultgeog.uu.se Phone: +46-(0)73-938-4911

ORCIDs: Marina Toger 0000-0003-4903-6971; John Östh 0000-0002-4536-9229; Ian Shuttleworth 0000-0003-0279-9103

**Author Contributions**

MT, IS, and JÖ designed research; MT, IS, and JÖ performed research; JÖ prepared data; MT, IS, and JÖ analysed data; MT, IS, and JÖ wrote the paper.

**Abstract**

Mobile phone data – with file sizes scaling into terabytes – easily overwhelm the computational capacity available to some researchers. Moreover, for ethical reasons, data access is often granted only to particular subsets, restricting analyses to cover single days, weeks, or geographical areas. Consequently, it is frequently impossible to set a particular analysis or event in its context and know how typical it is, compared to other days, weeks or months. This is important for academic referees questioning research on mobile phone data and for the analysts in deciding how to sample, how much data to process, and which events are anomalous. All these issues require an understanding of variability in Big Data to answer the question of how average is average? This paper provides a method, using a large mobile phone dataset, to answer these basic but necessary questions. We show that file size is a robust proxy for the activity level of phone users by profiling the temporal variability of the data at an hourly, daily and monthly level. We then apply time-series analysis to isolate temporal periodicity. Finally, we discuss confidence limits to anomalous events in the data. We recommend an analytical approach to mobile phone data selection which suggests that ideally data should be sampled across days, across working weeks, and across the year, to obtain a representative average. However, where this is impossible, the temporal variability is such that specific weekdays' data can provide a fair picture of other days in their general structure.

**Significance Statement**

Often mobile phone research requires selecting a sample of records because of computational or ethical restrictions and this raises questions of representativeness. Here we develop a methodology to make an informed sample choice by understanding the temporal variability inherent in the data. In COVID-19 research, for instance to measure how much human activity has changed, a comparative 'normal' baseline is necessary. But which days and hours are most 'normal' for this baseline? By analysing phone activity over a duration of more than 500 days using simple statistical procedures, we demonstrate a method to get the most stable findings by selecting 11.00 on Thursdays as the less variable day/hour – a day and hour with least variability than other weekdays.

**Introduction**

As the sun rises on any given day, people go to work or education, engage in other daily activities whilst yet others stay at home, collectively making up the vibrant buzz and spatial patterns of the society. Modelling this complex behaviour demands large amounts of data, feasible due to recent advances in data collection and processing. The advent of Big Data, whether from mobile phones (1), travel cards (2), or social media (3), enables mapping of the daily patterns to understand such phenomena as segregation (4) (5), spread of diseases such as Covid-19 (6) (7), human mobility (8) (9), and urban dynamics (10). The time-dimensional difference between traditional data-sources in the form of census data or population registers means that what was measured on decennial or annual periodicity now can be measured in real time. As a result, there is an increasing number of studies using finer temporal resolutions in societal studies but there are, to our knowledge, no studies indicating which days or hours that are representative for societal activities in general. There is also a growing interest in the ethical implications of the collection and use of Big Data and much discussion about how to visualise and analyse these large datasets (11) (12). We address these concerns by asking 'how normal is normal?'.
Understanding the periodicity and typical patterns in the data is important since one response to ethical/privacy concerns is to allow analysis only of a subset of the data and one quick answer to the large size of Big Data is similarly to subsample the data whether for a day, a week, a month or a particular place. These seemingly trivial questions are vital to our understanding of complexity and variability in Big Data and answering whether it is sufficient to analyse data for a day or a week to improve our understanding of societal processes such as segregation, whether all data should be analysed, or when and if there are decreasing returns to computing time in revealing the difference. Furthermore, to assess the impact of shock events such as extreme weather to get a better grasp on social resilience, it is important to study the natural dataset variability at different temporal scales and periodicities. After all, to understand how anomalous an event is, it is necessary to have a benchmark against which to judge it. The conceptual framework used to evaluate behaviour as revealed by our mobile phone dataset (supplementary Figure S1).

Previous studies looked at variability in mobile phone usage data, for instance (13) assumed weekly periodicity in CDR for selecting a benchmark in their study of emergency related human behaviour in a European country. Here we examine the data to determine the extent of weekly/monthly/hourly periodicity rather than assuming it is there. Variation in the number of users and in daily travel distances manifested weekly, monthly and seasonal regularities in population movement around the holidays in connection with the earthquake in Haiti, the analysis (14) requiring opening and processing the trajectories of individuals. Comparison (15) of the load volume on electrical infrastructure with phone activity in Senegal, so that phone activity can function as a proxy for level of development/electrification, also required processing of the files. A visual analytic approach to detecting anomalies in call activity in Senegal (16) entailed illustrating visually the temporal changes in weekends/religious holidays. These studies showed significant and detectable variability in mobile phone activity based on Call Detail Records (CDR). Our Network Detail Records (NDR) dataset comprises calls, SMS, MMS, data connections, and also silent handovers, thus having more frequent entries per user than the popularly used CDR. Given the structure of our dataset, the activity level is proportional to the number of observations/rows in the dataset and thus to the file size (see supplementary Figure S2). We propose a convenient estimate of variability in activity based on file sizes, without opening them, with a measurement of deviation between expected *vs* observed periodical temporal signature of mobile phone activity. Moreover, in our case mobile technology penetration in Sweden (17) is 98% so phone activity is likely to be representative of behaviour across all demographic groups.

**Results**

Two different time lags were used for the autocorrelation analysis for hours (Figure **1** a) and days (Figure **1** b). The repetitive pattern in day lag autocorrelation is easily detectable (Figure **1** b). The thin black line indicates that the seventh, fourteenth, twenty-first days (and so on) are more correlated than the days in between, pointing to weekday-specific file sizes.

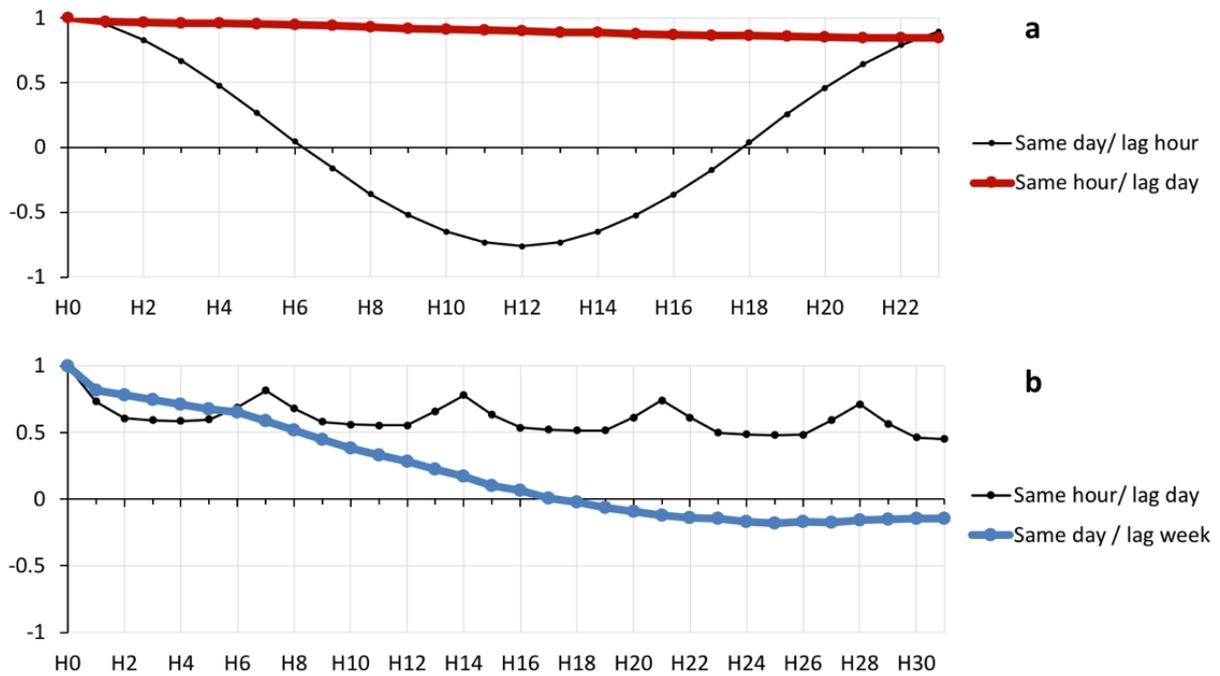

Figure 1 | Autocorrelation analysis: Two different time lags are illustrated for (a) hours or (b) days. In other words, correlations are conducted using the same hour (a) or the same day (b) and at a lag-distance that grows with a day's distance (for hours) or a week's distance (for days) for each step on the x-axis. The far-right part of the red lines is located at around 2/3 of a month distance (hours – a) and around 7 months' distance (days – b). The thin black lines depict a time lag that grows with one hour for each step on the x-axis (a) and a day for each step on the x-axis (b). Thus, the far right of the thin black line corresponds to a day's distance (hours – a) and to a months' distance (days – b).

The bold blue line stretches for a longer time span (lag distance is one week) and describes the annual repetitive pattern where the lowest correlation value for any specific weekday is found at lag of 26 weeks (six months) away. The strong repetitive hourly, daily, weekly and monthly patterns in the data indicate that file size can be used as a proxy for activity at the temporal scales illustrated here.

MLM is used to (1) estimate file size variance is a three-level hierarchical model and (2) estimate the impact of events and weather on levels of activity. **Table 1** shows the results of the empty model with no fixed effects and with standardised file size as the dependent variable. It shows that most of the variation is between hours (83.9%), followed by months (7.9%) and days (4.3%). An additional 3.9% of the variance cannot be explained but might be related other external factors such as weather and national events.

Table 1 | Empty model. The variation at each level, hour, day, and month-year, plus residual is saved to the estimate column. % variation indicates the share of variation at each level.

|  | *Coef.* | *Std. Err.* | *z* | *P>\|z\|* | *[95% Conf. Interval]* | |
|---|---|---|---|---|---|---|
| *Constant* | -0,0062 | 0,1874 | -0,03 | 0,973 | -0,3735 | 0,3610 |
| *Random-effects Parameters* | Estimate | Std. Err. | **% variation** | **% cumulative** | [95% | Conf. Interval] |
| *Variation (Hour):* | 0,8357 | 0,2432 | **83,9%** | **83,9%** | 0,4724 | 1,4784 |
| *Variation (Day):* | 0,0433 | 0,0057 | **4,3%** | **88,2%** | 0,0334 | 0,0560 |
| *Variation (MY):* | 0,0782 | 0,0024 | **7,9%** | **96,1%** | 0,0737 | 0,0831 |
| *Variation (Residual):* | 0,0390 | 0,0006 | **3,9%** | **100,0%** | 0,0379 | 0,0402 |
| *Number of observations =* | 11922 | | | | | |
| *LR test vs. Linear model: χ2=* | 31827,94 | | | | | |

Table **2** presents the results of the full MLM with fixed effects. The random effects parameters indicate that the hour-level variation drops to 81.1% while no change is detectable at day and month-year (MY) levels. The fixed effects explain around 2.3% (note that cumulative % is reaching 97.7%) of the variation. This may sound small, but is in line with the expectations since the majority of phone usage is connected to regular reoccurring events such as workdays and weekends, night-rest and active hours. Of the fixed effects it is clear that religious and secular holidays significantly and negatively reduce the file size (usage of phones) while major sports and TV media-events increased the file sizes. Hours with major transport breakdowns or out-of-the-ordinary weather have small effects although some are statistically significant.

Table 2 | Results from the full MLM with events and weather introduced as fixed effects

| FIXED EFFECTS | Coef. | Std. Err. | z | P>|z| | [95% Conf. Interval] | |
|---|---|---|---|---|---|---|
| EVENTS | | | | | | |
| Secular holiday | **-0,1220** | 0,0142 | -8,57 | **0,0000** | -0,1499 | -0,0941 |
| Religious holiday | **-0,1616** | 0,0128 | -12,63 | **0,0000** | -0,1867 | -0,1366 |
| Sports | **0,0316** | 0,0144 | 2,19 | **0,0290** | 0,0033 | 0,0599 |
| TV Media | **0,0349** | 0,0162 | 2,16 | **0,0310** | 0,0032 | 0,0666 |
| Weather & transport | 0,0032 | 0,0253 | 0,13 | 0,8990 | -0,0463 | 0,0528 |
| WEATHER | | | | | | |
| Air temp Malmö | 0,0015 | 0,0013 | 1,14 | 0,2560 | -0,0011 | 0,0040 |
| Air temp Stockholm | **0,0041** | 0,0015 | 2,73 | **0,0060** | 0,0011 | 0,0070 |
| precipitation Malmö | 0,0029 | 0,0057 | 0,51 | 0,6100 | -0,0083 | 0,0142 |
| Precipitation Stockholm | **0,0203** | 0,0076 | 2,66 | **0,0080** | 0,0053 | 0,0352 |
| precipitation Malmö squared | -0,0005 | 0,0014 | -0,37 | 0,7100 | -0,0034 | 0,0023 |
| Precipitation Stockholm Squared | -0,0004 | 0,0008 | -0,58 | 0,5590 | -0,0019 | 0,0010 |
| Air temp Malmö squared | **-0,0001** | 0,0001 | -2,19 | **0,0290** | -0,0002 | 0,0000 |
| Air temp Stockholm squared | **0,0002** | 0,0001 | 3,55 | **0,0000** | 0,0001 | 0,0004 |
| D precipitation Malmö | 0,0013 | 0,0063 | 0,21 | 0,8380 | -0,0110 | 0,0136 |
| D precipitation Stockholm | -0,0038 | 0,0073 | -0,52 | 0,6000 | -0,0180 | 0,0104 |
| D Air temp Malmö | 0,0025 | 0,0018 | 1,33 | 0,1830 | -0,0012 | 0,0061 |
| D Air temp Stockholm | 0,0028 | 0,0030 | 0,93 | 0,3500 | -0,0031 | 0,0088 |
| D precipitation Malmö squared | 0,0003 | 0,0009 | 0,31 | 0,7570 | -0,0015 | 0,0021 |
| D precipitation Stockholm squared | -0,0003 | 0,0005 | -0,57 | 0,5710 | -0,0014 | 0,0008 |
| D Air temp Malmö squared | 0,0000 | 0,0001 | -0,25 | 0,8000 | -0,0003 | 0,0002 |
| D Air temp Stockholm squared | **0,0043** | 0,0012 | 3,49 | **0,0000** | 0,0019 | 0,0068 |
| CONSTANT | -0,0686 | 0,1851 | -0,37 | 0,7110 | -0,4315 | 0,2942 |
| Random-effects Parameters | Estimate | Std. Err. | **% variation** | **% cumulative** | [95% | Conf. Interval] |
| Hour: | 0,8148 | 0,2372 | **81,8%** | 81,8% | 0,4605 | 1,4415 |
| Day: | 0,0425 | 0,0056 | **4,3%** | 86,1% | 0,0328 | 0,0550 |
| MY: | 0,0786 | 0,0024 | **7,9%** | 93,9% | 0,0740 | 0,0835 |
| Residual: | 0,0370 | 0,0005 | **3,7%** | 97,7% | 0,0359 | 0,0381 |
| Number of observations = | 11922 | | | | | |

In Table **3**, the third restricted model, an alternative empty model, is specified, which excludes festivals and national events leaving only 'normal' days. The results indicate that taking away unusual days strengthens the already clear temporality observed in the data; the variation attributable to hours increases to 84.2%, months to 8.0%, although that for days remains the same.

Table 3 | Restricted model: alternative empty model where event days were excluded

| | Coef. | Std. Err. | z | P>|z| | [95% Conf. Interval] | |
|---|---|---|---|---|---|---|
| **Constant** | **-0,00068** | 0,18816 | -0,000 | 0,997 | -0,3694 | 0,368115 |
| Random-effects Parameters | Estimate | Std. Err. | **% variation** | **% cumulative** | [95% | Conf. Interval] |
| Hour: | 0,8429 | 0,2453 | **84,2%** | 84,2% | 0,4765 | 1,4910 |
| Day: | 0,0429 | 0,0057 | **4,3%** | 88,5% | 0,0330 | 0,0556 |
| MY: | 0,0801 | 0,0025 | **8,0%** | 96,5% | 0,0753 | 0,0851 |
| Residual: | 0,0353 | 0,0006 | **3,5%** | 100,0% | 0,0342 | 0,0364 |
| Number of observations = | 10,715 | | | | | |
| LR test vs. Linear model: $\chi 2$= | 29034,35 | | | | | |

In the final analytical stage, the difference between observed file sizes and the values predicted by all three of the multilevel models for hours and days is analysed using RMSD. This shows which hours and which days are easier to predict. As a baseline, the three models produce RMSD values of ~0,1877 for the empty model, ~0,1834 for the full model and ~0,1737 for the restricted empty model. The variation is largest for the empty model, with slight reduction when introducing fixed effects (full model) and smallest when all days having events are removed from the analysis (restricted model).

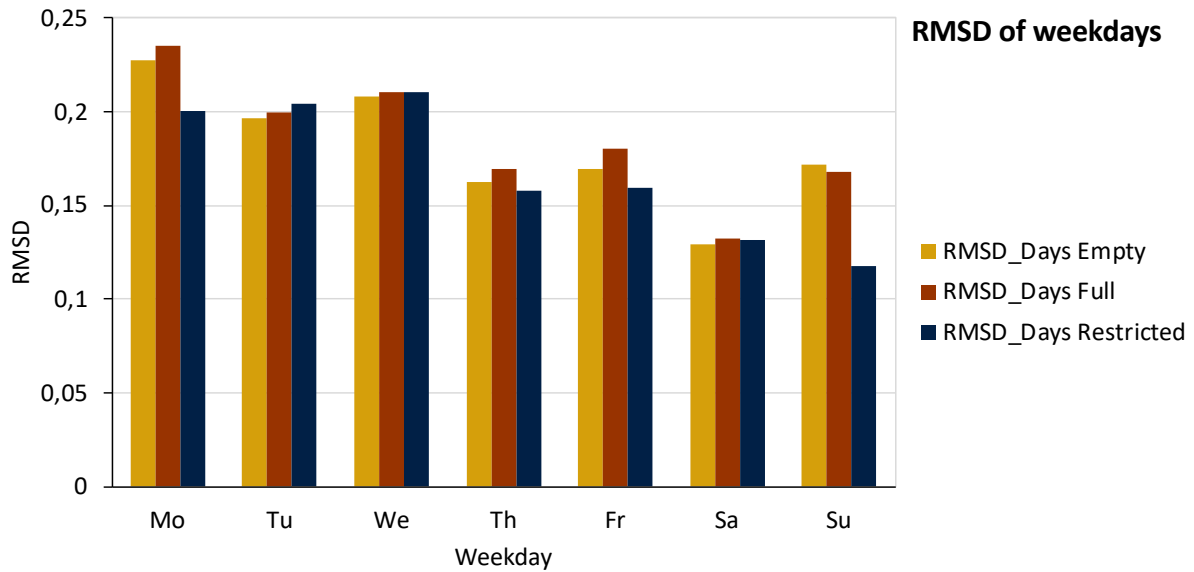

Figure 2 | RMSD by weekday for the three models: empty (yellow), full (red), and restricted (blue). Turns out Thursday is the most 'normal' weekday (and not Tuesday) and Saturday is the most normal weekend day.

Decomposing the files into weekdays (Figure **2**) pinpoints Mondays as days with the greatest RMSD value, indicating that it is difficult to predict file-size (i.e. phone usage varies more amongst Mondays than between other days). The opposite situation is found for Saturdays which tend to have low RMSD values. There is an interesting deviation between models for Sundays where all hours are kept in (empty and full) and the model where only days without events (such as holidays) are modelled (restricted model). See supplementary Figure S3 for graph of the relationship between standardized predicted and observed file size.

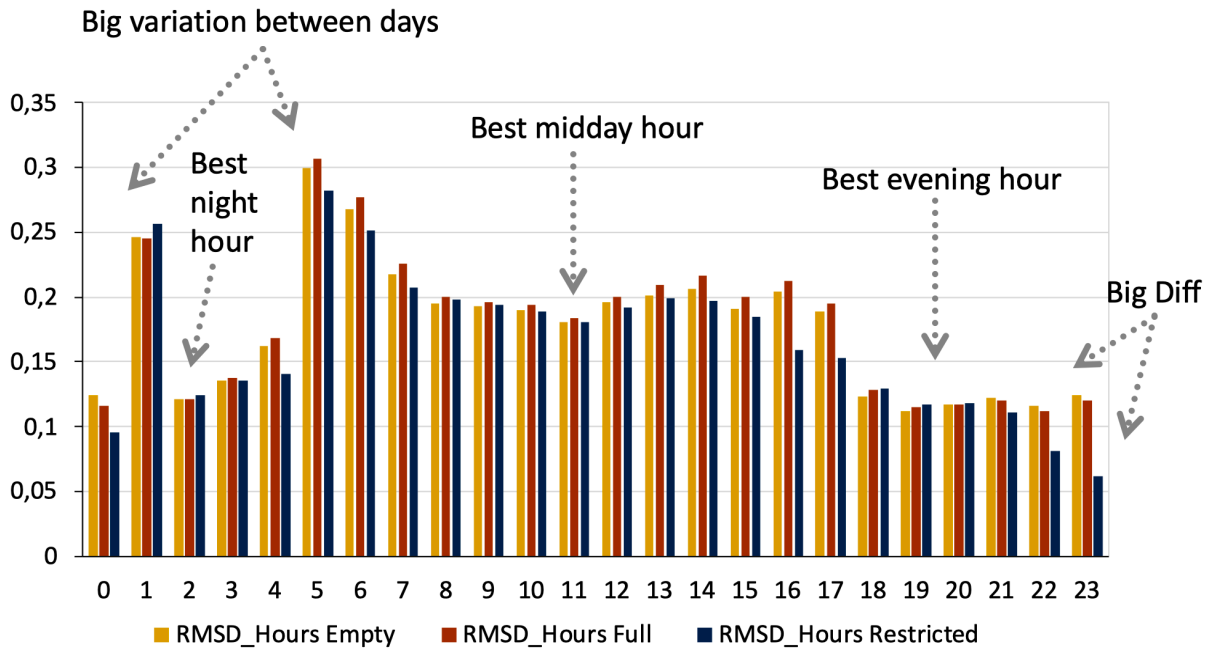

Figure 3 | RMSD by hour: empty (yellow), full (red) and restricted (blue) models. Some hours (e.g. 01:00-02:00) have high variation among days and thus should not be selected for subsamples.

The greatest RMSD for hour values (Figure **3**) are found at hours 1 (01:00-02:00), 5 and 6 (05:00-07:00). Early afternoon hours also have relatively high values. The least varying hour at night is 2 (02:00-03:00), daytime is 11 (11:00 12:00), and evening is between 19 and 20 (19:00-21:00). In the supplementary material (Figure S4), combined day-hour graphs are available where the RMSD for all hours across all days is shown.

**Discussion**

The mobile phone dataset analysed in the paper reveals temporal complexity in human behaviour at temporal scales of hours, days, weeks and months. The diurnal patterns shown are much as might be expected but the variability by day and month indicate the importance of considering these aspects. The method demonstrated to estimate the activity of mobile phone users from file size offers a quick and convenient way to extract patterns from datasets that are otherwise unwieldy thereby enabling generalisations to be made to set samples in their context. Our model (see Figure S3) closely predicts observed file sizes (and thus levels of activity and phone usage). It is noteworthy that the empty model performs as well in some regards as the full model which has explanatory terms for national holidays, events and extreme weather; this indicates the importance of hourly, daily and monthly temporal beats in the data, overlain with random variation, as the main drivers of activity. Swedes are therefore largely creatures of habit (but with a certain amount of random variation thrown in). Despite this, national holidays, other events, and weather do have some influence on behaviour although this is small compared to the rolling cycle of the day, week. month and year.

The daily and hourly patterns can be explained by how human activities are organised. The difficulty in predicting activity on Mondays and Fridays arises from their position at the start and end of the working week and their consequent variability across seasons and sensitivity to holidays and other events. Thus, mid-week days are more 'average', but Thursday performs better than Tuesday like one might expect. This also explains the volatility of the hours between 05:00 and 07:00 and 22:00 and 00:00 where behaviour will vary for the same reasons – other times of the day are less variable. The hourly variability therefore means most care should be taken in selecting which hours to sample; this is the largest source of variation in our data. However, month is also important and whereas neighbouring months are similar, the decrease in similarity as the temporal lag increases shows that if possible data should be extracted from more than one month, and ideally months half a year apart, if the objective is to capture the full variability of human behaviour.

There is a small number of high RMSDs. These are not correlated with any known extreme event. Closer inspection of these extreme file sizes (see supplementary Figures S5 and S6) which are more than 2 standard deviations from the series average indicates that these rare occurrences are usually in the early hours of the morning and during week days when file sizes are normally small. Their cause is unknown but can most probably be attributed to technical issues such as system updating and maintenance (the providers are reluctant to discuss these issues openly). It is important to note because these (and similar) technical issues may be present in other mobile phone (and other similar) big datasets. If these extreme cases are dropped as anomalies, then the model performs even better since there is less random variation. The decision whether to drop them or not would depend on the objective of the study. If the focus is on human behaviour, then data integrity related outliers should be dropped. However, if the data itself is under scrutiny, then it is open to discussion whether these cases should stay part of the analysis. Therefore, we kept them in.

It is likely that similar societies to Sweden will show the same temporal patterns of behaviour but that there will be increasing divergence as social difference increases by level of economic development, religion and world area (1). Therefore, these results cannot be generalised to all national contexts. However, the method of using mobile phone NDR file sizes as proxies for activity and human behaviour can be generalised assuming that data providers release the essential data of file size by hour, day, week and month so as to permit the analysis of variation as has been undertaken here.

**Materials and Methods**

**The phone data**

The data were obtained from a major Swedish mobile phone provider. At time of analysis the data series covers more than 500 days. Over time an archive of compressed data ranging on a byte-scale between terabytes ($10^{12}$) and petabytes ($10^{15}$) have been registered (the exact size cannot be revealed due to an agreement with the data provider). In a decompressed and analytically manageable format, the volume of the dataset becomes considerably bigger. The data are NDR which comprise calls, SMS and MMS messages, data uploads and downloads, and silent handovers. They record geographical movements at a small temporal scale of five-minute intervals nested within hours, days, weeks and months. As such, they provide a proxy for human activity and behaviour. For this study, the lowest five-minute level of the data is ignored as this requires processing of the files. Instead, we concentrate file size changes over hours, days, weeks and months (
Figure 4).

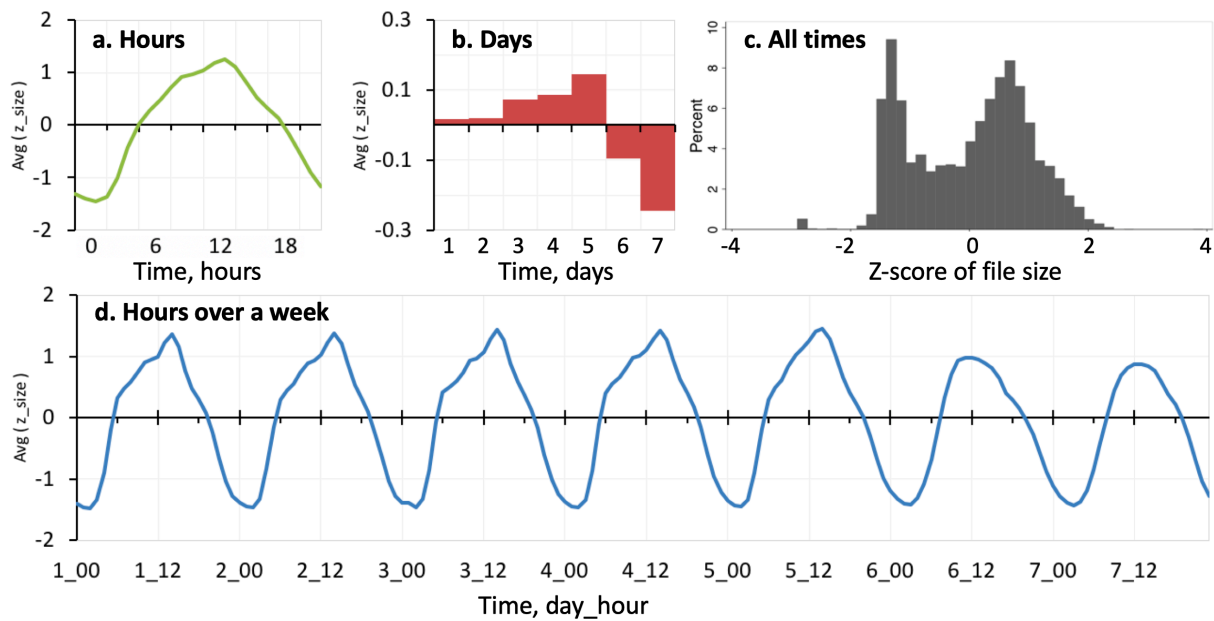

Figure 4 | Average standardized size of the hour-files: per hours (a) days (b), and over the week (d) and histogram of all file sizes (c).

No personal information about phone owners is known and the individual records are anonymised. The hourly files number 12,000 and rising. Due to ethical, legal and computational restrictions, often it is only possible to analyse a fraction of the data. Data subsamples are unpacked and transferred to a database called MIND; non-analysed data are registered but moved in their compressed format to a repository[1].
Total number of phone events (rows) is highly correlated with standardized file-sizes ($R^2$ = 0,9971, see supplementary Figure S1). The database is available in the supplementary material, including the standardized hourly file-size and all the explanatory variables.

**Weather data**

Precipitation and temperature are used to describe weather during the studied period. Both variables are available for download with a temporal resolution of hours which makes the weather data easy to integrate with MIND data. However, geography poses a problem. Since the MIND data is derived from users in all of Sweden the weather data should represent the weather situation valid for a large share of the individuals. The challenge is to generate representative weather data for a relatively big country with considerable differences in temperature and precipitation. We have chosen to draw data from Stockholm and Malmö which both are located in the populous south. The two cities are ranked number one and three in population size and are located close to the remaining bigger cities in Sweden. The online data are available from the Swedish meteorological surveys[2]. The two variables were temperature (in centigrade per hour, Air temp Malmö and Stockholm) and precipitation (mm per hour, precipitation Malmö and Stockholm). However, since also rate and magnitude of change may contribute to the model, additional variables have been created:

- squared precipitation per hour (precipitation Malmö and Stockholm Squared), and squared temperature per hour (Air temp Malmö and Stockholm squared), were added in case the relationship is not linear
- precipitation change per hour (D precipitation Malmö and Stockholm), temperature change per hour (D Air temp Malmö and Stockholm), was added to cater for reaction for a change in temperature (e.g. it got colder, so there are "dad, pick me up" calls, or it got warmer and people connect to suggest outdoor activities)
- squared precipitation change per hour (D precipitation Malmö and Stockholm squared) and finally, squared temperature change per hour (D Air temp Malmö and Stockholm squared),

---

[1] referred to as *NotOnOurMind*
[2] For information about weather stations see Supplementary materials.

which were added because people might be reacting to a size of change rather than its direction.

**Events data**

Various regular calendar events as well as unplanned events affect the behaviour of populations. Big enough events may affect the entire country contributing to changes in how people spend their time, their spatial and digital activity. In order to test if and potentially which events that affect the size of hour-files we have generated a set of dummy variables, coded 1 for the hours (or full days) they are observed, including the following: Secular holiday, religious holiday, sports, TV, media and, weather & transport (i.e. complete stop in traffic between main national nodes or extreme weather affecting a large proportion of the population)[3].

**Method**

The analysis relies on the relationship between activity, whether mobility or phone usage, as measured by the number of rows in a file and its size on disk. This avoids the daunting need to open every file to measure and count each activity. The robustness of this approach is demonstrated by high correlation between file-size and count of activities (supplementary Figure S2). The dependent variable for the analysis – aggregated to hours – is thus simply file size. Temporal patterns in the data are described and then analysed using established methodologies. **Autocorrelation across time** is used to assess hourly and daily time lags in activity. We conduct an autocorrelation test of the variation in file size by correlating the of hours (and days) files to lag hours (and days) files as specified in equation 1.

$$r_h = \frac{\sum_{t=1}^{N-h}(Y_t-\bar{Y})(Y_{t+h}-\bar{Y})}{\sum_{t=1}^{N}(Y_t-\bar{Y})^2} \quad \text{(eq.1.)}$$

The result $r_h$ is the correlation between any hour or day and time lag h that ranges between perfectly correlated, random or perfectly negatively correlated ($r_h$ values of 1, 0 and -1 respectively). By plotting the $r_h$ for a series of values where the time lag h is increasing by one hour (or one day for the day series) we depict changes in autocorrelation, as the distance in time lag increases, in the shape of a correlogram. The correlogram illustrates the trend-change in autocorrelation to test our hypothesis that the size of files follows a pattern that is strongly related to the repetitive nature of days and hours.

**Multilevel models (MLM)** are used because the dependent variable is sorted on time in a nested hierarchy of hours, days and months. MLM is used to estimate variance across temporal levels accounting for weather and special events. Temporal categories, enable detecting how much of the file-size variation can be assigned to hours, days or months but also see how much, and which temporal category of the variation can be explained by the fixed effects. The fixed effects are listed as event and weather dummy variables. Three different multi-level analyses are employed in this study. First, we run the multi-level regression (the empty model) using no explanatory variables, but with three temporal levels: hour, day and a combined month-year. The empty model is formulated as specified in equation 2.

$$y_{ijk} = \beta_{0ijk} + v_k + u_{jk} + e_{ijk} \quad \text{(eq.2.)}$$

Where $y_{ijk}$ refers to the file size z-score and subscripts $ijk$ represents hour, day and month-year levels respectively. The variation (or random errors) at different levels is expressed as: $v_k$ (month-year), $u_{jk}$ (day) and $e_{ijk}$ (hour). The empty model enables estimating the percentage of variation attributable to hours, days and month-years, but also enables assessing to what extent the introduction of fixed effects for events and weather in the full model reduce variation at different temporal levels.

The second and third multi-level analyses differ from each other in two ways: (1) the second (full) model makes use of all listed event and weather variables and (2) the third (restricted) model is an empty model (no fixed effects) but contains only the hours and days that are not associated with specific events. Therefore, the restricted model contains fewer cases. The full multi-level model can be expressed as specified in equation 3.

---

[3] List of events is included as supplementary material

$$y_{ijk} = \beta_{0ijk} + \beta_1 x_{1ijk} + \cdots + \beta_n x_{nijk} + v_k + u_{jk} + e_{ijk} \qquad (eq.3.)$$

Where $\beta_{0ijk}$ represents the intercept and $\beta_n x_{nijk}$ the time-specific regression coefficient between a fixed effect predictor and the dependent variable.

Predicted file-sizes from these MLM are compared with the observed using the **Root Mean Square Deviation** (RMSD) so as to measure the scale of temporal variation. By comparing the deviation between estimated and observed size of the hour-files, we trace which hours and weekdays tend to vary more or less from the average. By repeating the RMSD test for each hour, day and combined week-day and hour, we detect which hours and days are best suited for selection of a representative subsample. The RMSD can be expressed as in equation 4.

$$RMSD_t = \sqrt{\sum (Y_t - \widehat{Y}_t)^2 / n} \qquad (eq.4.)$$

Where $t$ represents specific time (hour or day), $Y_t$ and $\widehat{Y}_t$ - observed and estimated file-size, and $n$ - number of time units included in the analysis[4].

**Acknowledgments**

We thank the mobile phone data provider for the data. The work of MT and JÖ was performed with the support of Formas Research Council funding.

---

[4] Note that hour files are used also for day-analyses. In this case, the average observed size of hour-files on any given day is compared to the average estimated size.